\documentclass{article}
\usepackage{spconf,amsmath,graphicx}
\usepackage{amssymb}
\usepackage{url}
\usepackage{float}

\title{Generative Adversarial Source Separation}
\name{Y.Cem Subakan$^\flat$, Paris Smaragdis$^{\flat,\sharp}$\thanks{This work is supported by NSF grant \#1453104.}}
\address{$^\flat$UIUC, $^\sharp$Adobe Systems \\
  \{subakan2, paris\}@illinois.edu}
\begin{document}
%
\maketitle

\begin{abstract}
Generative source separation methods such as non-negative matrix factorization (NMF) or auto-encoders, rely on the assumption of an output probability density. Generative Adversarial Networks (GANs) can learn data distributions without needing a parametric assumption on the output density. We show on a speech source separation experiment that, a multi-layer perceptron trained with a Wasserstein-GAN formulation outperforms NMF, auto-encoders trained with maximum likelihood, and variational auto-encoders in terms of source to distortion ratio. 
\end{abstract}

\section{Introduction}
Many popular audio modeling/source separation algorithms such as Non-Negative Matrix Factorization (NMF) \cite{Smaragdis2003}, autoencoders \cite{Smaragdis2017}, or tensor factorization models \cite{Cemgil2011} are all generative models and they are trained with maximum likelihood (ML) which require the specification of output distributions. For instance, NMF models with different loss functions (e.g. KL-NMF, Euclidean NMF, IS-NMF) \cite{Fevotte2010}, actually have the same underlying mapping from latent space to observed space (same underlying network), but their performances typically differ on a given dataset. The output distribution/loss function therefore biases the model.

Generative Adversarial Networks (GANs) \cite{Goodfellow2014} offer a generative model learning framework, which does not require the specification of an output distribution. GANs are able to learn the processes which are implicitly defined via a transformation of a random variable. Namely, the generative process is defined such that a random latent variable is mapped to the data domain via getting transformed through a deterministic neural network. This removes the bias that comes from assuming a parametric output distribution and leads to more accurate modeling of distributions. \cite{Goodfellow2017} 


GANs have been very popular in computer vision since their first introduction \cite{Goodfellow2017}. However, to the best of our knowledge, usage of GANs in the audio modeling domain has been limited. In \cite{Pascual2017}, authors train de-noising networks by using an adversarial framework. In this paper, we propose using GANs to learn a generative model over magnitude spectrogram frames, which are used in a speech source separation task. 

Source separation is the task where the goal is to decompose a given signal into additive components which approximates the original sources as accurately as possible. In generative source separation, we train generative models to recover the sources from an observed mixture. We experimentally show on speech mixtures that an adversarially trained two layer perceptron outperforms NMF and ML-trained autoencoders in terms of source-to-distortion ratio \cite{Vincent2006bss-eval}. In our experiments, we have observed that the original GAN formulation in \cite{Goodfellow2014} is hard to train. We therefore showed the performance improvement over standard audio models with the more recent Wasserstein-GAN formulation \cite{Arjovsky2017}.


\section{Generative Supervised Source Separation} 
In generative source separation, an observed mixture signal $x \in \mathbb R^L$ is assumed to follow the generative process below:
\begin{align*}
  h_1 \sim p_{\text{latent}}(h_1), \; s_1|h_1 \;& \sim \; p_{\text{forward}_1}(s_1|h_1) \\
  h_2 \sim p_{\text{latent}}(h_2), \; s_2|h_2 \;& \sim \; p_{\text{forward}_2}(s_2|h_2) \\ 
  x|s_1, s_2 \;  \sim \;  &p_\text{mixture}(x | s_1 + s_2)
\end{align*}
where the source $s_k \in \mathbb R^L$, $k\in\{1, 2\}$, follow the distribution $p_{\text{source}_k}(s_k) = \int p_{\text{forward}}(s_k|h_k)p(h_k) d h_k$, where $h_k \in \mathbb R^K$ is a latent variable with lower dimensionality, such that $K < L$, and $p_{\text{forward}}(s_k|h_k)$ is the forward model for the sources. Given the sources, the mixture $x$ is assumed to be distributed according to the conditional distribution $p_\text{mixture}(x|s_1 + s_2)$, where $x$ is conditioned on the sum of the sources. Note that we have not yet assumed parametric forms for the distributions above. Also note that, in our experiments we consider the case where there are only two sources, although methods discussed can be generalized to more sources. To give a context on our audio application, mixture $x$ here corresponds to a column of a magnitude spectrogram.  

The goal in source separation is to compute accurate estimates for the sources given a mixture signal $x$. In supervised generative source separation, the approach is to first train the forward models $p_{\text{forward}_1}(.)$, $p_{\text{forward}_2}(.)$ such that the source distributions $p_{\text{source}_1}(.)$, $p_{\text{source}_2}(.)$ are approximated as best as possible. Given the trained models for both sources, in testing we compute source estimates $\widehat s_1$, $\widehat s_2$ such that the conditional distribution $p_\text{mixture}(x| \widehat{s_1} + \widehat{s_2})$ is maximized (or equivalently the reconstruction error for the mixture is minimized). In the next section, we describe the specifics on how to go through supervised source separation with maximum likelihood training.

\subsection{Maximum Likelihood Training for Sources}
A common way to go about approximating the source distributions is through assuming that the sources are generated by transforming a $K$ dimensional latent variable $h \sim p_{\text{latent}}(h)$, through a non-linear mapping (such as a neural network) $f_\theta(h)$ with parameters $\theta$, and adding noise to the transformed variable. This corresponds to the following generative model:
\begin{align}
  h \sim p_{\text{latent}}(h),\; s|h \sim  p_{\text{out}}(s; f_\theta(h)), 
\end{align}
where $p_{\text{out}}(.)$ is the output distribution which models the noise at the output of the mapping $f_\theta(h)$. E.g. When modeling spectrograms, $p_{\text{out}}(.)$ is usually taken as Poisson distribution, which corresponds to the unnormalized KL divergence. Under these modeling assumptions, the optimization problem for approximating source distribution is written as follows: 
\begin{align}
  \max_\theta \sum_t \log \int \mathcal{PO}(s^t; f_\theta(h)) p(h) dh, \label{eq:MLinttrain}
\end{align}
where the integral over the latent variable $h$ is intractable in the general case. Using variational auto-encoder framework in \cite{Kingma2013}, the objective in expression \eqref{eq:MLinttrain} can be maximized by computing a variational lower bound.

In practice however, especially in audio modeling, the integral over the latent variable $h$ is not computed, and only the conditional forward model $p_{\text{forward}}(s|h)$ is learnt, by simultaneously optimizing over the forward model parameters and the latent variables. This is written as the following optimization problem: 
\begin{align}
  \max_{\theta, h} \sum_t \log \mathcal{PO}(s^t; f_{\theta}(h)), \label{eq:MLtrain}
\end{align}
If $f_\theta(h)=Wh$, where $W, h \geq 0$, then this formulation corresponds to the widely used Non-Negative Matrix Factorization (NMF) model \cite{Smaragdis2003, Smaragdis2014-spm, Fevotte2010}. It is also possible to include the latent variable estimation part in the model with an auto-encoder. This results in the following optimization problem:  
\begin{align}
  \max_{\theta, \theta^{-1}} \sum_t \log \mathcal{PO}(s^t; f_{\theta}( f^{-1}_{\theta^{-1}}(s^t) )), \label{eq:MLtrain}
\end{align}
where $f^{-1}_{\theta^{-1}}(.) : \mathbb R^L \to \mathbb R^K$, is the encoder, and $f_{\theta}(.) : \mathbb R^K \to \mathbb R^L$ is decoder part. In \cite{Smaragdis2017}, $f_{\theta}(h) =\log (\exp( W_1 h) +1)$, and $f_{\theta}^{-1}(s) = \log(\exp( W_2 s) +1)$ is used.


The conceptual problem with the training objectives discussed in this section is that by picking a specific output distribution, we are sacrificing from the generality of the approximated source distributions. To remove this assumption, in this paper we use generative adversarial networks, which is a neural network framework for learning generative models without explicitly specifying an output distribution when training the generator network. 
\subsection{Adversarial Training for Sources} 
We have seen in the previous section that maximum likelihood training involves a parametric assumption for the output distribution. As an alternative, in this paper we propose using an implicit generative model in training, which does not require an explicit loss function. An implicit generative model for the sources can be specified as follows: 
\begin{align}
  h \sim p_{\text{latent}}(h), \; s = f_\theta(h), \label{eq:gangenerate}
\end{align}
where the source $s$ is deterministically related to the latent variable $h$, unlike the source model in the previous section. This process implies an intractable density function $p_{\text{model}}(.)$ for $s$ in the general case where $f_\theta(h)$ is a complicated non-linear mapping such as a neural network. Learning under implicit generative models is a currently a very active field of research \cite{Mohamed2017}. One way to attack this problem is to use discriminator function $D_{\xi}(.)$ which aims to distinguish between the samples generated from the model and the training instances. The goal in training is then becomes to generate samples using the process in expression \eqref{eq:gangenerate} so that, the discriminator $D_\xi(.)$ becomes unable to distinguish between the generated samples and the training data. This described setup is known as a generative adversarial network (GAN) \cite{Goodfellow2014}, and the corresponding minimax game is specified as follows: 
\begin{align}
  \min_\theta \max_\xi \mathbb E_s \log D_{\xi}(s) + \mathbb E_{h} \log (1 - D_{\xi}( f_{\theta}(h))),
\end{align}
This expression can be recognized as the sum of Bernoulli log-likelihoods, where $D_\xi(.)$ tries to maximize by outputting 1 for the training data $s$, and outputting 0 for the generated samples $f_\theta(h)$. The generator however tries to minimize the expression by \emph{fooling} the discriminator. It can be shown that under some assumptions this scheme minimizes the Jensen-Shannon divergence between the actual source distribution $p_\text{source}(.)$ and the model distribution $p_{\text{model}}(.)$. However in practice, this scheme is unstable, and usually suffers from the mode collapse problem where the learnt distribution $p_{\text{model}}(.)$ only captures a subset of the actual sample space. \cite{Goodfellow2017}. This is unfortunately not acceptable for our source separation application. 

An alternate formulation known as the Wasserstein-GAN, alleviates the mode collapse problem by minimizing the Wasserstein-1 distance between the learnt and data distributions, which results in smooth gradients \cite{Arjovsky2017}. Authors show that this can be achieved with following minimax game: 
\begin{align}
  \min_\theta \max_{\xi\in \mathcal W } \mathbb E_s D_{\xi}(s) - \mathbb E_{h} D_{\xi}( f_{\theta}(h)),
\end{align}
where $\mathcal W$ denotes the set for parameters $\xi$ for which $D_\xi(.)$ will be $\gamma$-Lipschitz continuous, for some $\gamma$. In the algorithm provided in the paper, this constraint is achieved by clipping the weights $\xi$. In our experiments, Wasserstein GANs showed significant improvement over the original GAN formulation. Note that $D_\xi(.)$ is referred to as critic in this formulation.  


\subsection{Testing}
\label{sec:testing}
After training the forward models, given an observed mixture, the estimates $\widehat s_1$, $\widehat s_2$ for the sources is obtained by minimizing the reconstruction error via finding the optimal latent variables as inputs to the forward models: 
\begin{align}
  \widehat h_1,\; \widehat h_2 \; = \;\arg \max_{h_1, h_2} \log p_{\text{mixture}}(x ; f_{\widehat \theta_1}(h_1) + f_{\widehat \theta_2}(h_2) ), 
\end{align}
and then we get the estimates for the sources by setting $\widehat s_1$, $\widehat s_2$ = $f_{\widehat \theta_1}(\widehat h_1)$, $f_{\widehat \theta_2}(\widehat h_2)$, where $\widehat \theta_1$, $\widehat \theta_2$ denote the trained network parameters.  

An extra benefit we get by training our generative models with GANs is that, in addition to the generator networks $f_{\widehat \theta_k}(.)$, we also get discriminators/critics $D_{\widehat \xi_k}(.)$. We can therefore use them in the separation stage to score how much the obtained source looks like the instances in training set. We also noticed that using a smoothing term to enforce smooth first difference across time improves the quality of the estimated sources for both GANs and Maximum likelihood based auto-encoders. Therefore, the optimization for separating the sources, given $T$ mixture spectrogram columns $x^{1:T}$, becomes the following: 
\begin{align}
  \max_{h_1^{1:T}, h_2^{1:T}} &\frac{1}{T}\sum_{t=1}^T \log p_{\text{mixture}}(x^t ; f_{\widehat \theta_1}(h_1^t) + f_{\widehat \theta_2}(h_2^t) ) \notag \\
  & + \frac{\alpha}{T}\sum_{t=1}^T \left ( D_{\widehat \xi_1} (f_{\widehat \theta_1}(h_1^t) ) + D_{\widehat \xi_2} (f_{\widehat \theta_2}(h_2^t) ) \right ) \notag \\ 
  & + \frac{\beta}{T-1}\sum_{t=1}^{T-1} \left(\sum_{k=1}^2 \| f_{\widehat \theta_1}(h_k^{t+1}) -f_{\widehat \theta_1}(h_k^{t}) \|_1  \right ), \label{eq:gantest}
\end{align}
where $\alpha$ is a trade-off scalar between the reconstruction quality and discriminator/critic score. In our experiments we fixed $\alpha=0.1$, but it can also potentially be optimized on a validation set. For the smoothing term, we used $\beta=0.1$. Finally, note that for magnitude spectrograms it is very common to use Poisson distribution for $p_{\text{mixture}}(.)$.

\section{Empirical Results} 
To show the validity of using GANs in source separation, we compare adversarially trained networks with auto-encoders trained with maximum likelihood, variational auto-encoders and NMF. 

The experiment set-up is as follows: We form mixtures of male and female speaker utterances, and corresponding training data from the TIMIT speech corpus \cite{timit}. To form the training/test data pairs, we randomly pick male and female speaker pairs from the \emph{train} folder of the TIMIT corpus. Each speaker has 10 available utterances. For both speakers, of the 10 available utterances, we use 9 for training and 1 for testing. The resulting training set for each source is around 30 seconds long. The selected test utterances are around 3 seconds, and the mixture signal is obtained by mixing the test utterances at 0 dB. We form 25 such mixtures/training sets and test each algorithm on these randomly selected sets (The speaker pairs are the same across algorithms). As the preprocessing step, we compute Fourier spectrograms of the utterances. We use 1024 point FFT, and a hop size of 256. The learning and source separation are performed on the columns (Fourier magnitude vectors for each time window) of the magnitude spectrograms. When reconstructing the separated sources in the time domain, we use the Wiener filtering equation:
\begin{align}
  \widehat s_k^\text{time} = \text{ISTFT}(\frac{\widehat s_k}{\widehat s_1 + \widehat s_2} \odot x \odot x^\text{phase}), \; \text{for} \; k \in \{1,2\}, 
\end{align}
where $x$ and $x^\text{phase}$ are respectively the magnitude and phase spectrograms of the mixture. The magnitude spectra for the estimated sources are denoted by $\widehat s_1$ and $\widehat s_2$. The estimated time domain signal is denoted by $\widehat s_k^\text{time}$. The division and the multiplication $\odot$ are both element wise, and ISTFT(.) designates the inverse short time Fourier transform operation to get the time domain signal from a complex Fourier spectrogram.   
\begin{figure*}[tb]
   \centering
   \includegraphics[scale=0.5]{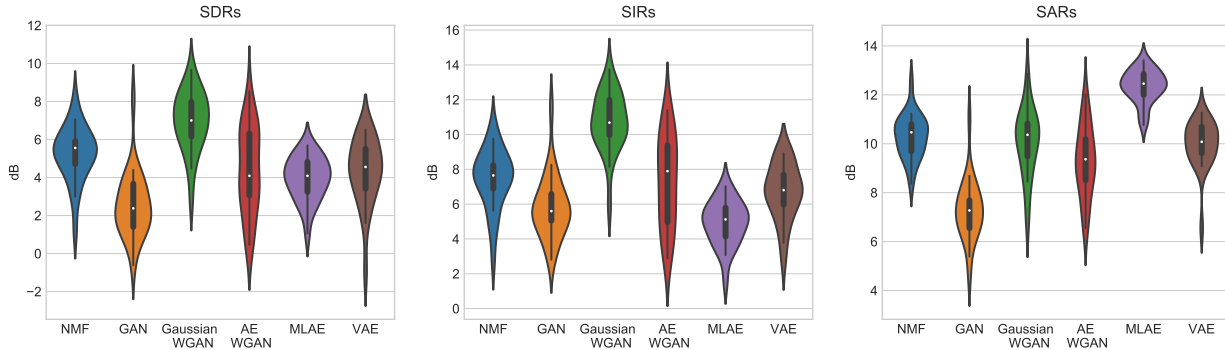}
   \caption{The distributions of the BSS eval scores for our speech source separation experiment. Acronyms for each algorithm are indicated below each violin plot. In order from left to right the algorithms are ordered as: NMF, Standard GAN, Wasserstein GAN with Gaussian inputs, Autoencoding Wasserstein GAN, Autoencoder trained with maximum likelihood, and Variational Autoencoder. Each violin shows the distribution of the corresponding score in dB over different speaker pairs. The subplots are organized as follows: \textbf{Left:} SDR scores, \textbf{Middle:} SIR scores, \textbf{Right:} SAR scores. }
   \label{fig:exp}
\end{figure*}
We obtain results on the following models: 
\begin{itemize}
\setlength{\itemsep}{1pt}
  \setlength{\parskip}{0pt}
  \setlength{\parsep}{0pt}
  \item KL-NMF model. 
  \item The auto-encoder model suggested in \cite{Smaragdis2017}, trained with maximum likelihood using a Poisson likelihood (equivalently unnormalized KL divergence). 
  \item Standard GAN with Gaussian random inputs. 
  \item Wasserstein GAN with Gaussian random inputs. 
  \item Autoencoding Wasserstein GAN, where instead of Gaussian random inputs, we feed the training samples to the generator network. 
  \item Variational Autoencoder with Gaussian prior on the latent variable, as in \cite{Kingma2013}, and Poisson likelihood at the output. 
\end{itemize}
For all GANs, we used the following architecture for the generator:
\begin{align}
  f_\theta(h) = \text{SP}( W_2 \text{SP}(W_1 h)), \label{eq:gennet}
\end{align}
where $\text{SP}(.)$ is the soft-plus nonlinearity, such that $\text{SP}(x)=\log(\exp(x) +1)$. Note that we have omitted the bias terms from the equation to reduce clutter.
For all GANs with Gaussian random inputs, we used 513 dimensional inputs $h$ (This is the dimensionality of the data items since we use 1024 point fft), and 100 hidden units. Therefore $W_1$ was of size $100\times 513$, and $W_2$ was of size $513\times 100$. For the auto-encoding GAN, and the auto-encoder trained with maximum likelihood, the network architecture of generator/forward model are exactly the same, except that the inputs are the data items $s$, instead of random variable $h$. For VAE, we used the encoder $f^{-1}_{\theta^{-1}}(s)= W_2 \text{ReLU}(W_1 s)$, both for the mean and the variance terms of the latent variable, where $W_1$ was of size $100 \times 513$, and $W_2$ was of size $20 \times 100$. For encoder of VAE, we used $f_\theta(h) = \text{SP}(W_3 h)$, where $W_3$ was of size $513 \times 20$. 

For the discriminator/critic networks of GANs, we use the following architecture: 
\begin{align}
  D_\xi(s) = \sigma( V_2 \tanh( V_1 s)), 
\end{align}
where $V_1$ is of size $90\times513$ (we use 90 hidden units for the discriminator), and $V_2$ is of size $1 \times 90$. In standard GANs, $\sigma(.)$ is the sigmoid function. In Wasserstein GAN, we do not use a non-linearity at the end of the network, and therefore $\sigma(.)$ is the identity function. This gives smoother gradients.   

In training and testing, for all neural network models we use the RMSprop algorithm \cite{RMSprop} with a learning rate of $0.001$. During the training of GANs, we do 5 iterations of discriminator/critic updates per generator update. For all neural network models, we do 4000 training iterations, and 20000 test iterations. For Wasserstein-GAN we clip the critic parameters at $-0.01$ and $0.01$ for lower and upper limits respectively. 

We report the BSS-eval \cite{Vincent2006bss-eval} scores obtained after recovering the sources from the mixture signals. The BSS-eval scores are Source to Distortion Ratio (SDR), Source to Interference Ratio (SIR), and Source to Artifacts Ratio (SAR), where SDR being the summary measure on how good the separation is. For each speaker pair, we have averaged the BSS-eval scores of the recovered sources, and in Figure \ref{fig:exp}, with violin plots we show the distribution of the averages of the two BSS-eval scores over all speaker pairs.

Experiments indicate that, The Wasserstein GAN with a Gaussian noise input outperforms NMF, ML auto-encoder and Variational auto-encoder in terms of source to distortion ratio. Note that we are obtaining these results with very similar underlying networks. For all models except VAE, we have kept the exact generator architecture defined in Equation \eqref{eq:gennet}. 
We have also observed that the standard GAN formulation is not very reliable. Although occasionally we have seen good SDRs with it, we have observed through inspecting its outputs that it is not able to capture the variety in the source spectrogram distribution as good as the Wasserstein GAN, and therefore the source separation performance of the standard GAN is not as good. We have also experimented with training an auto-encoder with adversarial training, and have seen that although it is less reliable than the Wasserstein GAN with Gaussian inputs, it is sometimes able to give great SDRs. In general, adversarial methods give great SIRs, by losing a bit from SAR, especially compared to the ML-autoencoder. Finally, note that the code for our experiments is available at \url{https://github.com/ycemsubakan/sourceseparation_misc}.
\section{Conclusions}
In this paper we have experimentally shown that Wasserstein GANs can obtain good performance in generative source separation. In addition to not requiring the specification of an output distribution, GANs fit into the source separation task nicely since the discriminator/critic functions help in source separation. We believe that there exists many research opportunities to use GANs in the audio domain. One natural next step from this paper is to extend the results showed in this paper with an end-to-end generative adversarial audio model.

\bibliographystyle{IEEEbib}
\bibliography{refs17}

\end{document}